\documentclass[twocolumn,twoside,slac_two]{revtex4}
\usepackage{graphicx}
\usepackage{fancyhdr}
\usepackage{ref}
\begin{document}

\title{Pulsar observations with the \emph{Fermi} LAT: what we have seen}

%

\author{L. Guillemot\footnote{E-mail: guillemo@mpifr-bonn.mpg.de}, on behalf of the \emph{Fermi} LAT Collaboration}

\affiliation{Max-Planck-Institute f\"ur Radioastronomie, Auf dem H\"ugel 69, 53121 Bonn, Germany}

\begin{abstract}
A year after \emph{Fermi} was launched, the number of known gamma-ray pulsars has increased dramatically. For the first time, a sizable population of pulsars has been discovered in gamma-ray data alone. For the first time, millisecond pulsars have been confirmed as powerful sources of gamma-ray emission, and a whole population of these objects is seen with the LAT. The remaining gamma-ray pulsars are young pulsars, discovered via an efficient collaboration with radio and X-ray telescopes. It is now clear that a large fraction of the nearby energetic pulsars are gamma-ray emitters, whose luminosity grows with the spin-down energy loss rate. Many previously unidentified EGRET sources turn out to be pulsars. Many of the detected pulsars are found to be powering pulsar wind nebulae, and some are associated with TeV sources. The \emph{Fermi} LAT is expected to detect more pulsars in gamma rays in the coming years, while multi-wavelength follow ups should detect \emph{Fermi}-discovered pulsars. The data already revealed that gamma-ray pulsars generally emit fan-like beams sweeping over a large fraction of the sky and produced in the outer magnetosphere.\\

\end{abstract}

\maketitle

\thispagestyle{fancy}


\section{Introduction}

After the \emph{Compton Gamma Ray Observatory} (CGRO) ended its activity in 2000, seven rotation-powered pulsars were known to emit in gamma rays, all normal pulsars, plus a couple of low-significance detections, including the millisecond pulsar PSR J0218+4232 (see \cite{Thompson2004} for a review of CGRO results on pulsars). More recently, the AGILE telescope \cite{Tavani2009} reported the detection of two other normal pulsars in gamma rays \cite{Pellizzoni2009,Halpern2008} plus a few other marginal detections, including that of another millisecond pulsar \cite{Pellizzoni2009}, yet to be confirmed. Less than ten pulsars were hence known to emit in gamma rays at the beginning of the $\emph{Fermi}$ mission, a small number compared to the almost 2000 known at radio wavelengths. 

Because of this limited sample, all attempts to model the fundamental phenomena producing high-energy emission in pulsar magnetospheres were so far confronted by a severe lack of observational data.  Namely, it was unclear whether high-energy photons are produced above the magnetic poles, or rather in gaps formed at higher altitude in the magnetosphere (see \cite{Harding2007} for a review of pulsar gamma-ray emission models). These models predict different features in pulsar light curves (signal intensity as a function of the rotational phase): number of gamma-ray peaks if any, phase separation between peaks if more than one, and phase separation between components at different wavelengths. Emission spectra also differ from one model to one another: for normal pulsars emission close to the surface would suffer magnetic pair attenuation, leading to a super-exponential energy cutoff in spectra, while for higher altitude emission the expected energy cutoff would be simple-exponential. Because they assume different emission geometries, the models predict different ratios of radio-loud and radio-silent gamma-ray pulsars, as their radio beam may or may not sweep across the Earth. Before $\emph{Fermi}$, only one such radio-silent gamma-ray pulsar had been detected, the Geminga pulsar, so that this ratio remained largely unconstrained. Finally, it was unclear whether millisecond pulsars (MSPs) could emit high-energy emission, and if they do, how their emission would compare to that of normal gamma-ray pulsars. The prospects for detections of MSPs were high though, after the few marginal detections in EGRET and AGILE data. The launch of the \emph{Fermi} observatory on 11 June 2008 provided a new opportunity to study the high-energy emission from pulsars and address the above questions.

The Large Area Telescope (LAT) \cite{Atwood2009}, main instrument of the \emph{Fermi} observatory, is sensitive to photons between 20 MeV and more than 300 GeV. It has an unprecedented sensitivity, thanks to its large field of view of $\sim$ 2.4 sr, large effective area ($\sim$ 8000 cm$^2$ above 1 GeV on axis) and improved angular resolution (0.6$^\circ$ at 1 GeV). It also has an excellent timing accuracy, better than 1 $\mu$s \cite{Onorbitcalib}, crucial for pulsar observations. Furthermore, the survey observation strategy of the $\emph{Fermi}$ observatory makes the LAT a prime instrument for the discovery of new sources. We review here the results of pulsar observations with the \emph{Fermi} LAT during its first year of activity, and discuss the vast prospects following these results.

\section{\emph{Fermi} LAT observations of pulsars}

\subsection{Observations of known pulsars}

There are two different approaches for the observation of pulsars with \emph{Fermi}. One is to search for gamma-ray pulsations from known pulsars, by phase-folding the photon arrival dates using pulsar rotation parameters measured at other wavelengths, namely radio and X-rays. Because radio and X-ray telescopes cannot monitor all $\sim$ 2000 known pulsars, a list of $\sim$ 230 prime candidates for detection in gamma rays has been defined prior to the launch, including the seven CGRO detections \cite{Smith2008}. These prime targets were then monitored by large radio telescopes around the world, as well as X-ray telescopes, aiming at measuring their rotation parameters (pulsar ephemerides) and keeping them accurate over the \emph{Fermi} observation range. Parameters were also obtained for \emph{a priori} secondary targets, expanding the number of pulsars to be searched for pulsations with the LAT to more than 700 objects.

\begin{figure}
\includegraphics[scale=0.4]{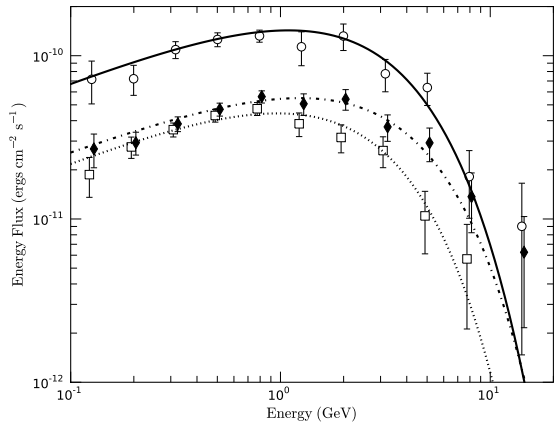}
\caption{Spectral energy distribution for the PSR J2021+3651 pulsar. The solid line shows the total flux emitted by the pulsar. The dotted and dot-dash lines correspond to the first and second peak in the gamma-ray light curve. See \cite{FermiJ2021} for additional details on this Figure.\label{J2021}}
\end{figure}

Using these parameters, the LAT detected gamma-ray pulsations from a number of normal radio-loud pulsars, all highly energetic. Two of these pulsars, J0659+1414 \cite{Fermi6PSRs} and J1048$-$5832 \cite{FermiJ1048_J2229}, had been seen with EGRET with low significance. Other objects lie in the error box of EGRET unidentified sources \cite{Hartman1999}: J1028$-$5819 in 3EG J1027$-$5817 \cite{FermiJ1028}, J1420$-$6048 in  3EG J1420$-$6038 \cite{Fermi6PSRs}, J2021+3651, also seen with the AGILE telescope \cite{Halpern2008}, in 3EG J2021+3716 \cite{FermiJ2021} and J2229+6114, in 3EG J2227+6122 \cite{FermiJ1048_J2229}. The \emph{Fermi} LAT has so far detected a total of 22 of these objects with high-confidence, over 100 MeV \cite{FermiPSRCat}. Table 3 of \cite{FermiPSRCat} gives the gamma-ray pulse shape parameters for these pulsars. Although the nature of a few light curves are unclear because of limited photon statistics, the typical light curve is two-peaked, with a separation between these peaks of $\sim$ 0.4, and the first gamma-ray peak lagging the main radio component by 0.1 to 0.2. Good examples of this typical shape are e.g. the Vela pulsar \cite{FermiVela} or PSR J2021+3651 \cite{FermiJ2021}. A few exceptions to this trend exist though, like the one-peaked PSR J2229+6114 \cite{FermiJ1048_J2229}, for which the unique gamma-ray peak lags the radio component by approximately half a rotation. The emission spectra have been examined as well (see Table 5 of \cite{FermiPSRCat} for a summary of spectral results). As illustrated in Figure \ref{J2021} for PSR J2021+3651, the energy spectra are found to be generally well fit by exponentially cutoff power laws, with cutoff energies below 10 GeV, in agreement with what had been measured with the EGRET telescope \cite{Thompson2004}. 

\begin{figure}
\includegraphics[scale=0.45]{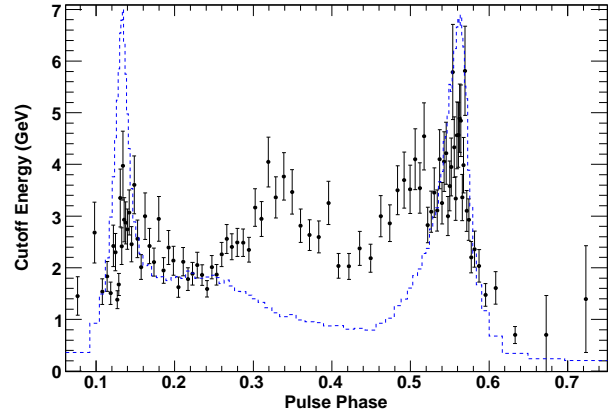}
\caption{Exponential cutoff energy as a function of the rotational phase, for the Vela pulsar. Figure adapted from \cite{TJJ}.\label{vela_ecvsphase}}
\end{figure}

Among radio-loud normal pulsars, the ones detected by the EGRET telescope or other previous experiments are in general prime targets for detailed spectral analyses because of their brightness, although the LAT has seen other bright pulsars that EGRET would have detected if accurate rotational parameters were available at that time, such as PSR J2021+3651. Early in the $\emph{Fermi}$ mission, the high-resolution light curves produced with LAT data for the Vela pulsar revealed a small third peak between the two main gamma-ray peaks, shifting to larger phases with increasing energy \cite{FermiVela}. The large amount of gamma-ray photons accumulated during this first year of data taking now enables very fine analysis of its emission properties as a function of rotational phase. Figure \ref{vela_ecvsphase} shows the energy cutoff of the emitted spectra as a function of phase. The important variations of the cutoff energy between the two main peaks is consistent with the above description of a third peak shifting towards higher rotational phase with increasing energy. Similarly complex behaviors of the spectral index and the cutoff energy are observed for other bright EGRET pulsars such as Geminga, J1057$-$5226, J1709$-$4429, or J1952+3252 \cite{OC,MR}, providing valuable input for the theoretical models.

\begin{figure*}
\includegraphics[scale=0.28]{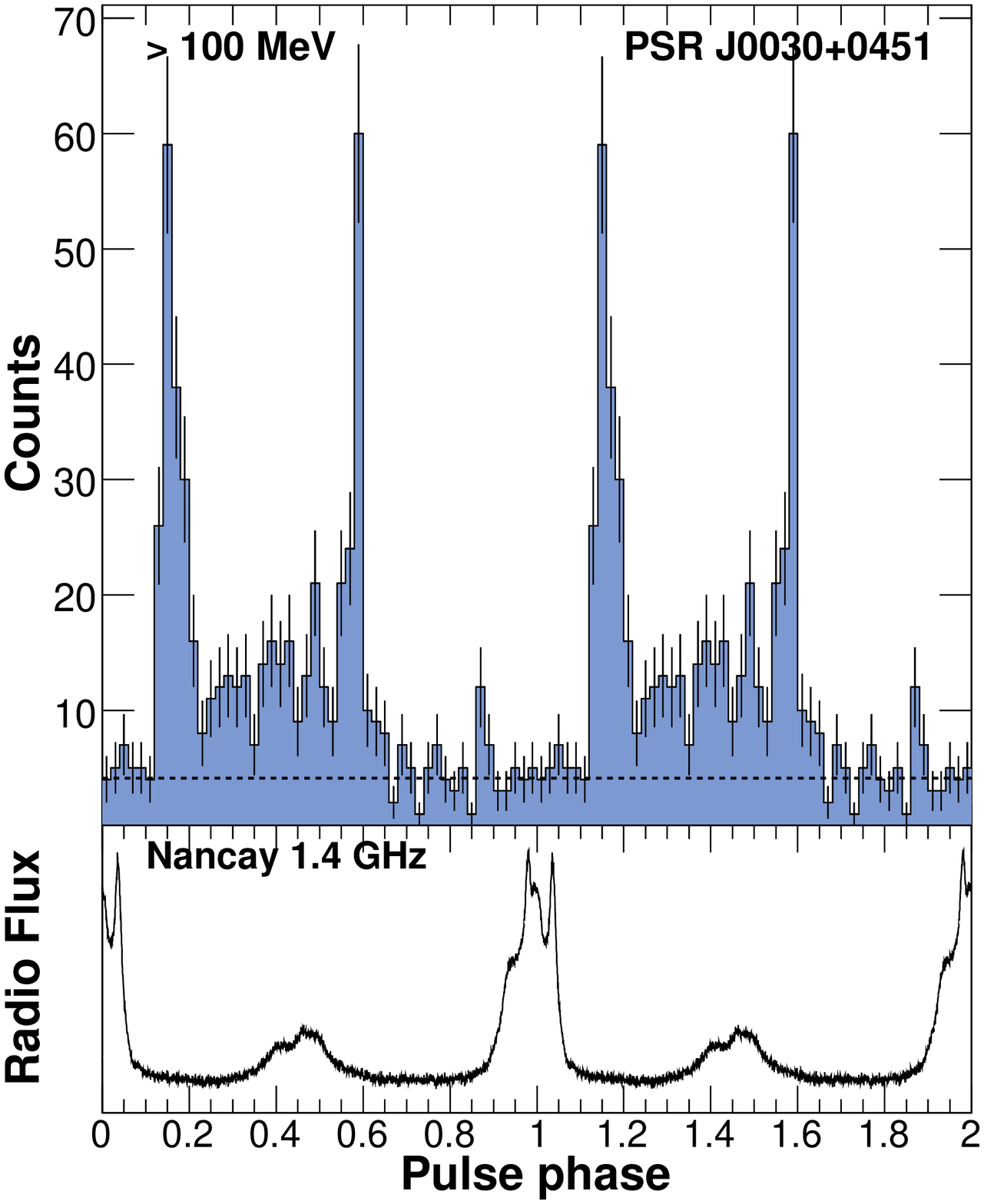}
\includegraphics[scale=0.28]{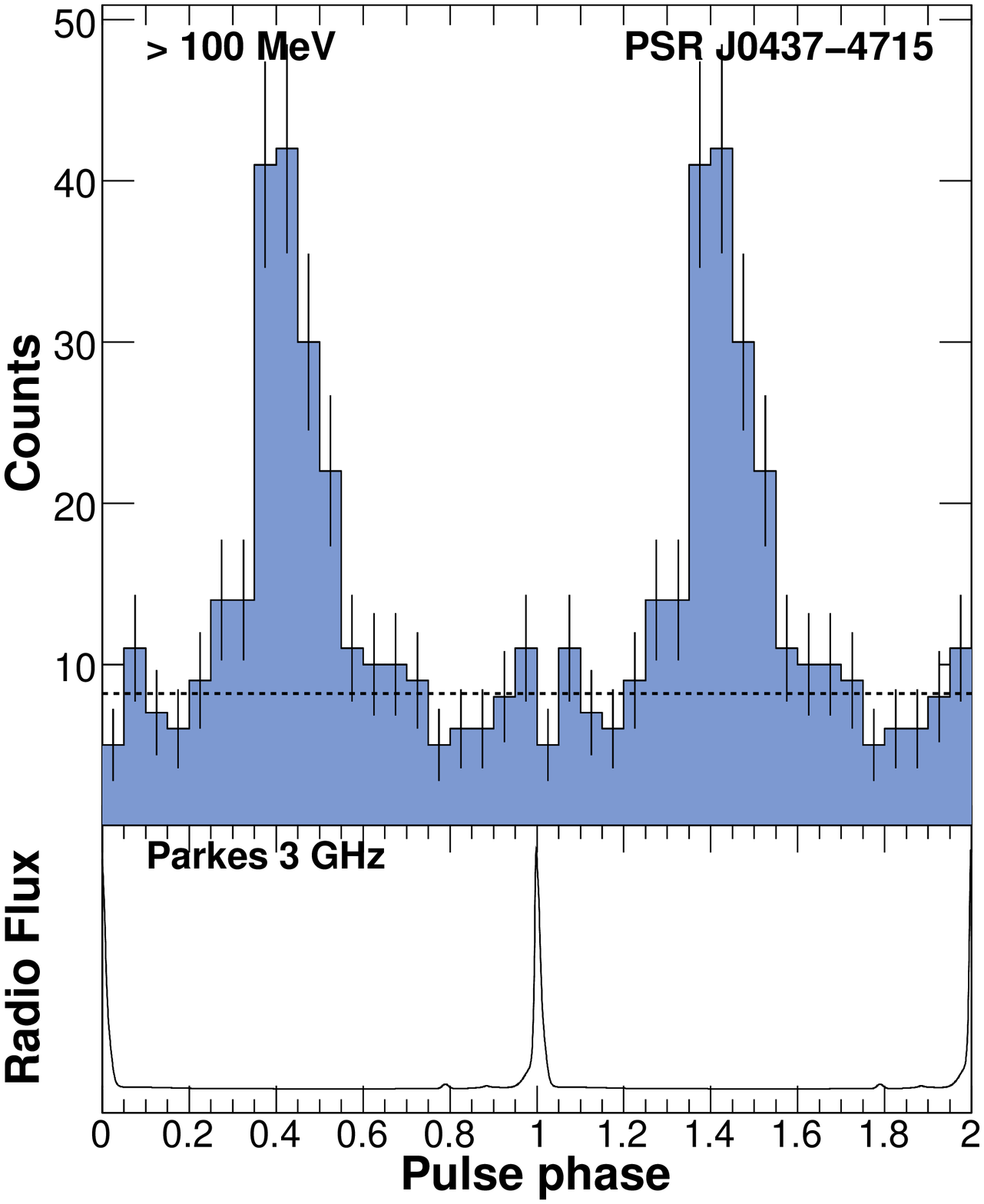}
\includegraphics[scale=0.28]{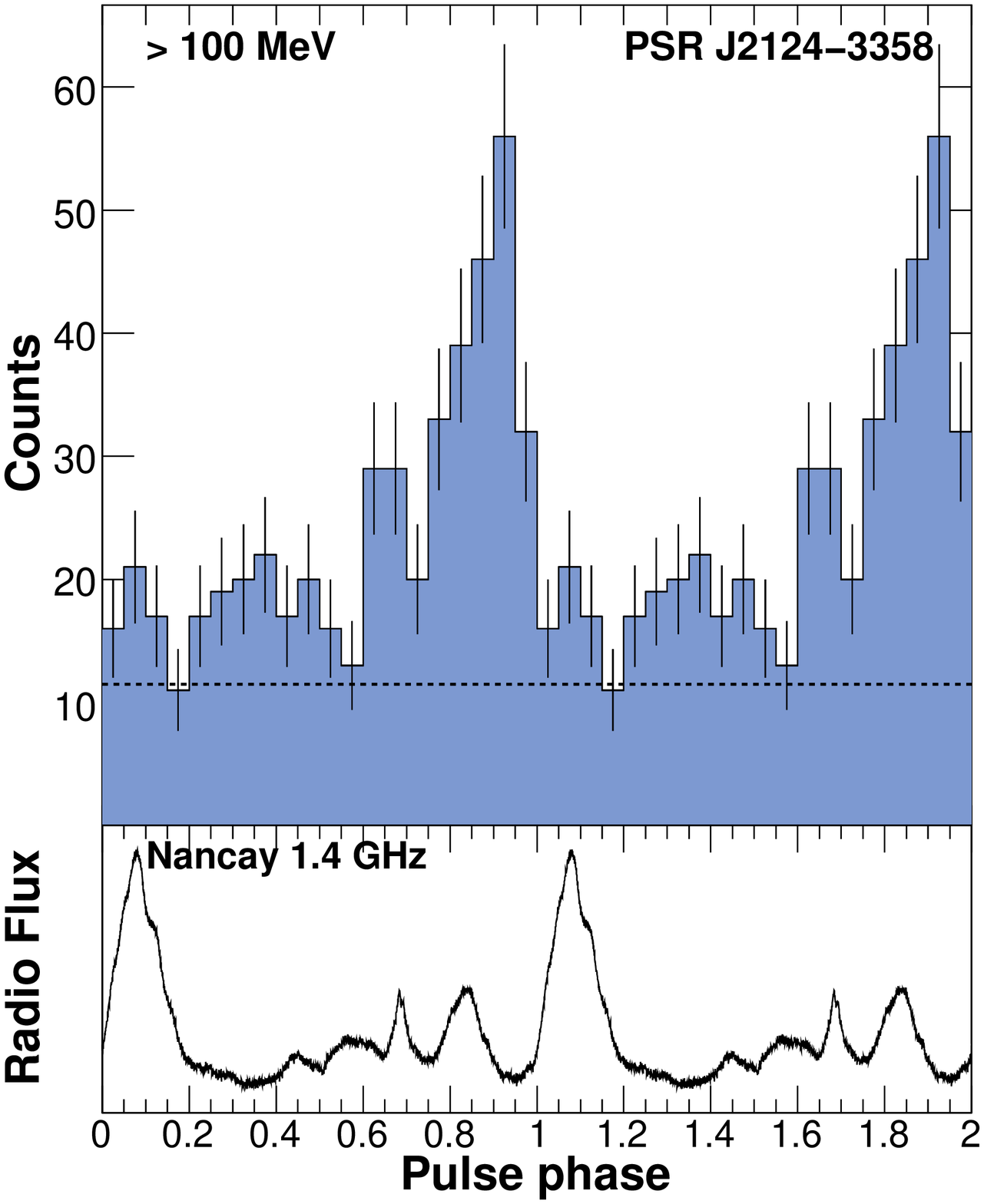}
\caption{Radio (lower panels) and gamma-ray (upper panels) pulse profiles for three of the nine millisecond pulsars detected so far with the \emph{Fermi} LAT. Two rotations are shown for clarity. Additional details on the construction of these light curves can be found in \cite{FermiMSPs}.\label{phasosMSPs}}
\end{figure*}

The search for pulsed gamma-ray emission from MSPs has been fruitful as well: pulsations were found for the nearby isolated MSP PSR J0030+0451 after three months of data taking, making this pulsar the first MSP firmly detected in gamma rays \cite{FermiJ0030}. Similarly to normal gamma-ray pulsars detected with the LAT, J0030+0451 lies in the error box of EGR J0028+0457, listed in the revised EGRET catalog of gamma-ray sources \cite{Casandjian2008}. After nine months of observation, the LAT detected pulsations from eight galactic MSPs, including the confirmation of PSR J0218+4232 as a gamma-ray pulsar \cite{FermiMSPs,Kuiper2000}. Figure \ref{phasosMSPs} shows radio and gamma-ray light curves for a selection of the eight detected MSPs. The light curves seen so far for MSPs resemble those of normal pulsars, as illustrated in Figure \ref{phasosMSPs}: PSR J0030+0451 has two peaks in gamma rays, separated by $\sim$ 0.45 and its first gamma-ray peak lags the main radio peak by $\sim$ 0.15, making its pulsar shape comparable to that of Vela or J2021+3651. PSR J0437$-$4715 shows one broad gamma ray peak, lagging the radio peak by $\sim$ 0.4, similar to what is observed for the normal pulsar J2229+6114. PSR J2124$-$3358 seems to exhibit a more complex light curve, as does the radio emission, with a main broad peak and possible secondary features. More data will help determine the peak multiplicity as is also the case for the normal pulsar J1420$-$6048 \cite{FermiPSRCat}. Similarly, their spectral properties do not differ fundamentally from that of normal pulsars (see Table 1 of \cite{FermiMSPs}); although super-exponentially cutoffs could not be tested, because of the limited size of the gamma-ray sample. As the LAT continues to accumulate photons, this newly revealed population of gamma-ray MSPs grows. For instance, the LAT recently discovered faint pulsed emission from the binary MSP PSR J0034$-$0534 \cite{FermiJ0034}. 

\subsection{Detections of new pulsars}

\begin{figure*}
\includegraphics[angle=270,scale=0.63]{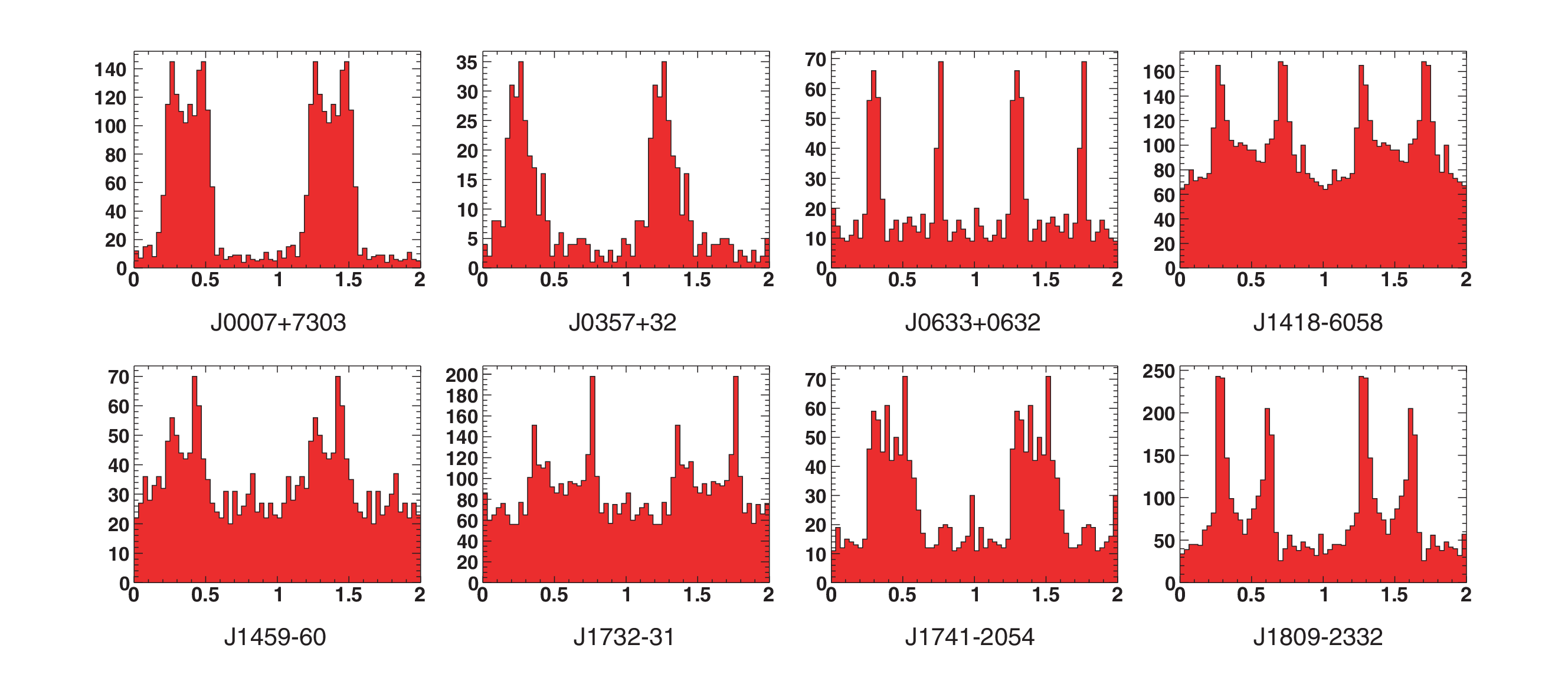}
\caption{Gamma-ray pulse profiles for a subset of the pulsars discovered with the \emph{Fermi} LAT. Two pulsar rotations are shown for clarity. Details on the construction of these profiles can be found in \cite{FermiBlind}.\label{BlindLC}}
\end{figure*}

The other approach is to search the gamma-ray data for yet unknown pulsars. The gamma rays are rare though: the LAT typically gets a few hundred photons for several months of data taking on a particular source. Because of these long integration times, direct Fourier transforms of the signal are computationally intensive. One can however search for gamma-ray pulsations in differences between photon arrival times. This so-called ``time-differencing'' technique dramatically reduces the needed computational power \cite{Atwood2006,Ziegler2008}. Although this technique did not foster new detections of gamma-ray pulsars in the EGRET data, it has been extremely efficient with \emph{Fermi} data: 16 slowly-rotating pulsars were found after five months of gamma-ray observation, including a pulsar in the young supernova remnant CTA1, another pulsar coincident with the isolated neutron star RX J1836+5925, and other objects likely associated with supernova remnants or pulsar wind nebulae \cite{FermiCTA1,FermiBlind}. In total, 13 of these 16 LAT-detected pulsars lie in the error box of third EGRET catalogue unidentified sources. Figure \ref{BlindLC} shows a sample of the observed gamma-ray light curves for these 16 pulsars. As expected, these light curves resemble those of other gamma-ray pulsars: they are generally two-peaked, with a few showing one broad peak. Their spectral properties are also consistent with that of the gamma-ray population (see Table 5 of \cite{FermiPSRCat}). More recently, a search for pulsars in 10 months of data using $\sim$650 LAT source positions unassociated with possible active galactic nuclei yielded the detection of eight new objects \cite{FermiBlind2}, expanding the number of LAT detected pulsars to 24 objects. Among these eight new pulsars, only one might be associated with an EGRET unidentified source, indicating that the \emph{Fermi} LAT is perhaps finishing the identification of unknown pulsars hidden behind EGRET sources, after one year of observation.

\begin{figure}
\includegraphics[scale=0.4]{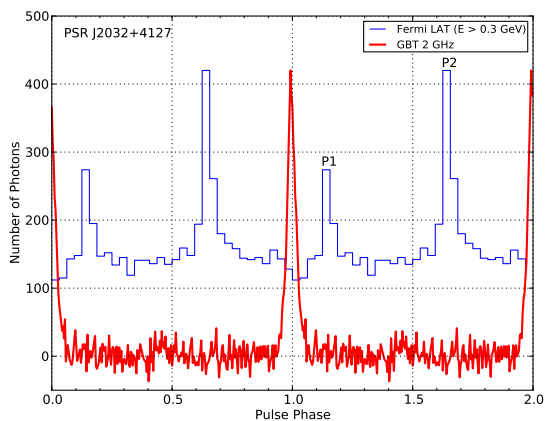}
\caption{Gamma-ray light curve of PSR J2032+4127 along with the radio profile measured with the GBT radio telescope \cite{Camilo2009}. Two other LAT-detected pulsars have so far been detected at radio wavelengths, at the Green Bank and Arecibo telescopes.\label{J2032LC}}
\end{figure}

As discussed above, the ratio of radio-loud to radio-quiet gamma-ray pulsars is an important discriminator for theoretical models. The LAT discovered pulsars are not necessarily radio-quiet, and thus need to be searched deeply for radio pulsations, helped by the pulsar positions and rotational parameters derived from the gamma-ray observations. So far, radio pulsations have been detected for three of them: PSR J1741$-$2054 and J2032+4127 at the Green Bank Telescope (GBT) \cite{Camilo2009} and PSR J1907+0602 at the Arecibo telescope \cite{FermiJ1907}. Figure \ref{J2032LC} shows the two-peaked gamma-ray light curve of PSR J2032+4127 as observed with the LAT, and the phase-aligned radio emission profile, providing the radio to gamma-ray phase separation, again very similar to that of other radio-loud pulsars. The other GBT-detected pulsar, PSR J1741$-$2054, is faint and close, and may have the lowest radio luminosity of all known radio pulsars. Similarly, the J1907+0602 detected with the Arecibo telescope is extremely faint ($\sim$ 4 $\mu$Jy flux density at 1.4 GHz), although the distance inferred from the dispersion measure is much larger than that of J1741$-$2054. The search for radio counterparts to the LAT-detected pulsars is still ongoing, the small number of radio detections reported so far nevertheless indicates that a large fraction of LAT-detected pulsars are radio-quiet.

\section{What do we learn?}

To summarize the \emph{Fermi} LAT observations of pulsars during its first year of operation, \textbf{a total of 55 pulsars} have so far been detected in gamma rays with high confidence. These include:

\begin{itemize}
\item \textbf{24 previously unknown normal pulsars}, discovered in the LAT data. Among these 24 gamma-ray pulsars, pulsations at radio wavelengths have been detected for three. 
\item \textbf{31 radio and X-ray loud pulsars}, detected in gamma rays using rotational parameters measured at radio telescopes. This number includes \textbf{22 normal pulsars} (among which are the six EGRET pulsars) and \textbf{9 millisecond pulsars}. 
\end{itemize}

For both normal and millisecond pulsars, most gamma-ray light curves have two sharp peaks, while a few objects have one broad peak. The gamma-ray emission beam then seems to consist of a fan-like beam. Depending on the viewing geometry, this beam can either intersect the line of sight of the observer twice, resulting in two sharp peaks, or graze the line of sight, resulting in a broad structure. This emission geometry matches well the prediction of the Outer Gap (OG) and Slot Gap (SG) models, for which the gamma-ray emission is produced at high altitude in the magnetosphere. In these models, gamma-ray pulsars emit high-energy photons in fan-like beams, that cover large fractions of the sky. The evidence for high-altitude emission in OG or SG scenarios are supported by spectral measurements. The emitted spectra are consistent with exponentially cutoff power laws, with cutoff energies below 10 GeV in all cases. As mentioned above, this spectral shape reveals the absence of magnetic pair attenuation, expected in case of emission produced at low altitude, where the magnetic field is most intense. Thus, both light curve and spectral properties favor outer magnetospheric emission for gamma-ray pulsars.

\begin{figure}
\includegraphics[scale=1]{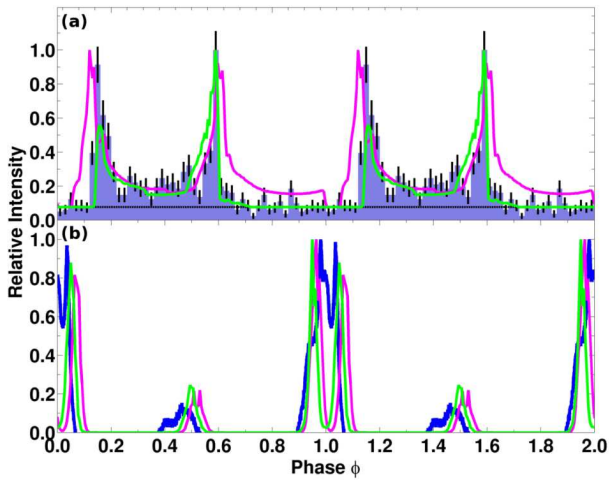}
\caption{Gamma-ray (upper panel) and radio (lower panel) light curves for the J0030+0451 millisecond pulsar, fitted with the Outer Gap (green) and Slot Gap (magenta) emission models. The observed emission is well reproduced by the two high-altitude emission models. Figure taken from \cite{Venter2009}.\label{0030Venter}}
\end{figure}

The observed similarities between normal and millisecond pulsars in terms of gamma-ray emission indicate that the same mechanisms are operating for both pulsar classes. Indeed, the Outer Gap and Slot Gap emission scenarios fit the observed light curves of gamma-ray millisecond pulsars well in most cases \cite{Venter2009,Venter2009_2}. Figure \ref{0030Venter} shows the observed radio and gamma-ray light curve for PSR J0030+0451, and fits obtained with the Outer Gap and Slot Gap emission models. These fits provide valuable constraints on the pulsar geometry (magnetic axis and line of sight inclination angles relative to the rotation axis, $\alpha$ and $\zeta$), generally in agreement with independent measurements, when available. Interestingly though, the PSR J1744$-$1134 and J2124$-$3358 MSPs are exclusively fit by the Pair Starved Polar Cap (PSPC) model, for which the gamma-ray emission is produced above the polar caps, at high altitude in the magnetosphere (see \cite{Venter2009}). High-altitude emission is thus favored for both millisecond and normal pulsars, although the location of the particle acceleration site may vary.

\begin{figure*}
\includegraphics[scale=0.67]{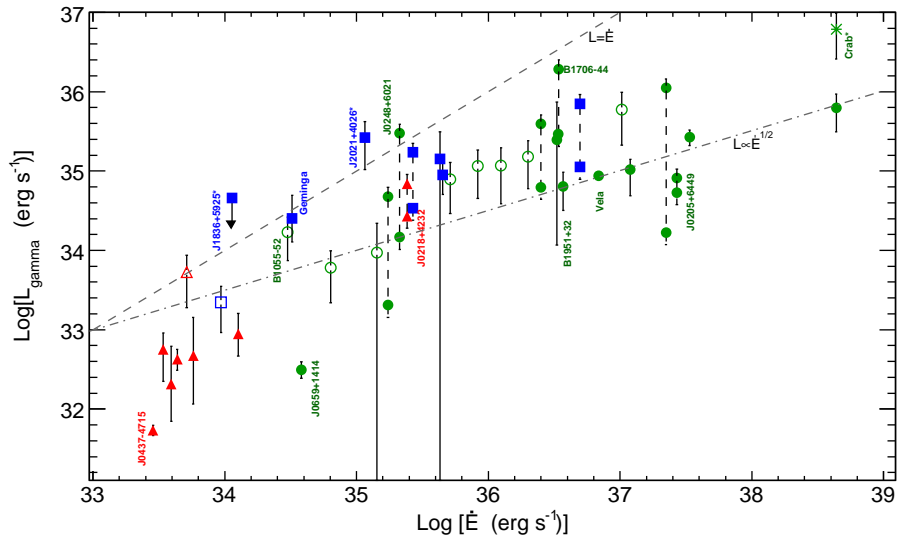}
\caption{Gamma-ray luminosity $L_\gamma$ above 100 MeV as a function of the spin-down energy loss rate $\dot E$, for gamma-ray pulsars with known distances. Pulsars with two distance estimates have two markers connected with dashed error bars. Red triangles show the gamma-ray millisecond pulsars, blue squares show the LAT-discovered pulsars and green circles are normal radio-loud pulsars. Pulsars whose distance is based on the dispersion measure only are shown with open symbols. Additional details on this plot can be found in \cite{FermiPSRCat}.\label{LvsEdot}}
\end{figure*}

After the observations of six pulsars in gamma rays with the EGRET telescope, there was evidence that the gamma-ray luminosity $L_\gamma$ was proportional to $\sqrt{\dot E}$, where $\dot E$ is the rotational energy loss rate via electromagnetic braking \cite{Thompson2004}. The question of the conversion of spin-down energy loss rate into gamma-ray luminosity was therefore a key question to be addressed with the LAT. Knowing the integral energy flux measured at the telescope, $h$, and the pulsar distance $d$, one can calculate the gamma-ray luminosity $L_\gamma$, given by: 

\begin{equation}\label{luminosity}
L_\gamma = 4 \pi f_\Omega h d^2
\end{equation}

In this expression, $f_\Omega$ is a flux correction factor, which depends on the pulsar configuration ($\alpha$ and $\zeta$ angles) and the assumed beaming geometry \cite{Watters2009}. This factor reflects the fact that the flux emitted towards the Earth may be different than that averaged over the sky, and is close to 1 for normal pulsars in outer magnetospheric emission models, with important variations though (see Table 1 of \cite{Watters2009}). This factor is also shown to be close to 1 for gamma-ray millisecond pulsars \cite{Venter2009}. Figure \ref{LvsEdot} shows the gamma-ray luminosity for pulsars with known distances, assuming an $f_\Omega$ of 1 for all pulsars. The luminosity values suffer important uncertainties, because of badly known distances or emission geometries. Nevertheless, the gamma-ray luminosity trend seen with EGRET for high $\dot E$ pulsars seems to stand. For the low $\dot E$ objects, including MSPs for which precise parallax distance measurements are generally available, the gamma-ray luminosity more seems to correlate with $\dot E$ rather than with $\sqrt{\dot E}$, indicating an inflection in the luminosity versus spin-down rate relation. Several objects in this low $\dot E$ regime are however found to be above the $L_\gamma = \dot E$ line, implying an absurd efficiency of conversion $\eta = L_\gamma / \dot E$ greater than 100\%. Their distances are probably overestimated. More precise distance estimates, and more gamma-ray pulsars populating the 10$^{34}$ to 10$^{35}$ erg/s region of this plot might help see where the inflection occurs.

\begin{figure*}
\includegraphics[scale=0.8]{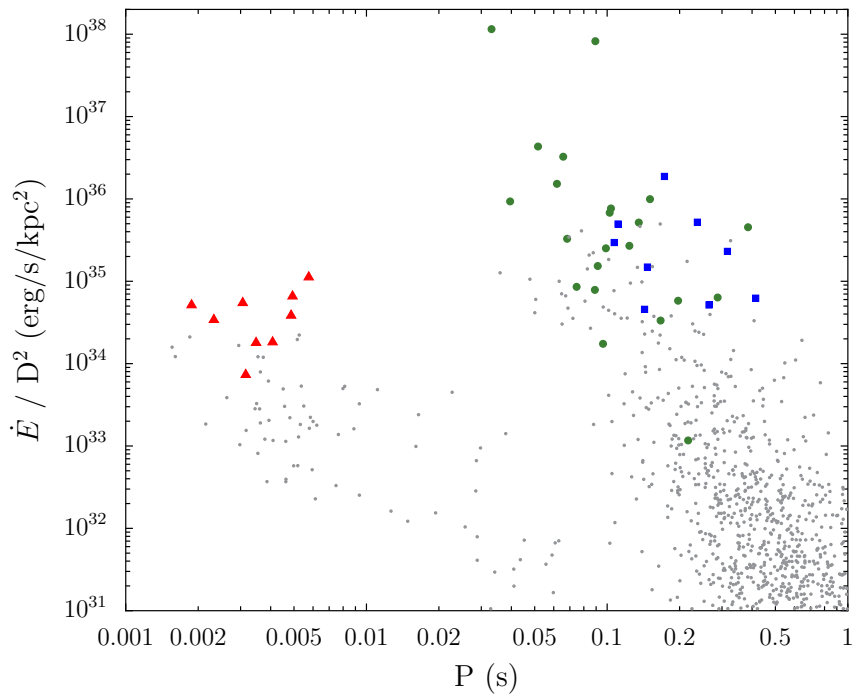}
\caption{``Spin-down flux'' $\dot E / d^2$ as a function of period for pulsars with known distances. Red triangles show the gamma-ray millisecond pulsars, blue squares show the LAT-discovered pulsars (accurate distances are unavailable for most of these pulsars, so that only a few of them are represented) and green circles are normal radio-loud objects. Gray dots are currently undetected pulsars. For these objects, DM distances were assumed, using the NE2001 galactic electron density model \cite{NE2001}. When possible, $\dot E$ values have been corrected for the Shklovskii effect \cite{Shklovskii} which artificially increases the apparent spin-down rate $\dot P$ of pulsars with important transverse velocity. Globular cluster pulsars are not represented in this plot.\label{Edotd2_P0}}
\end{figure*}

As most known pulsars have spin-down energy loss rates below 10$^{35}$ erg/s, it is reasonable to assume the $L_\gamma \propto \dot E$ relationship and thus use $\dot E / d^2$ as a criterion for pulsar visibility; although it obviously lacks the \emph{a priori} unknown beaming geometry information. Figure \ref{Edotd2_P0} shows the ``spin-down flux'' $\dot E / d^2$ as a function of period for pulsars with known distances. As expected, the detected pulsars lie in the upper part of the plot. The lowest MSP on this graph is PSR J1614$-$2230, whose distance is clearly overestimated, leading to an gamma-ray efficiency of more than 100\%, as can be seen in Figure \ref{LvsEdot}. The kinematic Shklovskii contribution \cite{Shklovskii}, which makes the apparent spin-down rate $\dot P$ larger than its intrinsic value, is therefore overestimated as well, and the actual $\dot E$ value must then be larger than we think it is. With both a smaller distance and a larger $\dot E$, this pulsar should be higher in the $\dot E / d^2$ versus $P$ plot. Bearing this in mind, we see that nearly all MSPs above a certain $\dot E / d^2$ threshold have been detected with the \emph{Fermi} LAT. Similarly, a good fraction of high $\dot E / d^2$ normal pulsars are detected. Also, most LAT-detected pulsars are not shown in this plot because of unknown distances, and they are expected to populate the upper part of the plot. Most detected normal and millisecond pulsars are found to be above a common $\dot E / d^2$ threshold, again indicating similarities in gamma-ray emission mechanisms. 

There are a number of possible reasons for the existence of gray dots in the upper part of Figure \ref{Edotd2_P0}: some pulsars may be faint and therefore need additional LAT data to be detectable (for instance, low galactic latitude pulsars, currently undetectable because of the intense galactic diffuse background). Accumulated photon counts will doubtless increase the number of detected gamma-ray pulsars. Some high $\dot E / d^2$ pulsars may be bright gamma-ray pulsars, for which beams do not point towards the Earth. Additionally, some pulsars may well not emit in gamma rays at all. Detailed geometrical studies of individual objects might help understand the causes of non-detection. Also, it should be emphasized that DM-based distances are uncertain and some undetected high $\dot E / d^2$ pulsars may be more distant, explaining their gamma-mutism. 

\begin{figure*}
\includegraphics[scale=0.8]{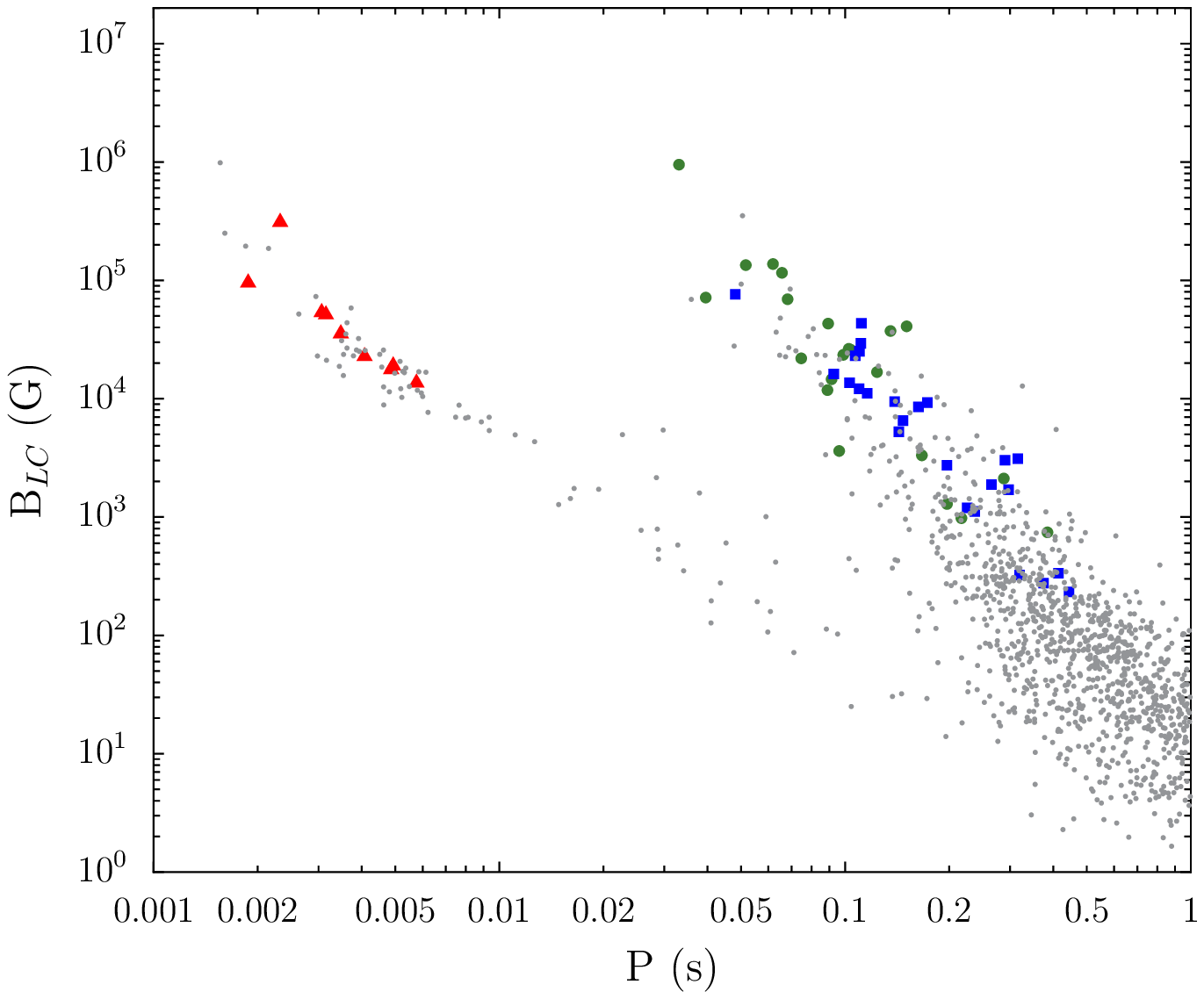}
\caption{Magnetic field at the light cylinder $B_{LC}$ as a function of the rotational period $P$ for known pulsars. The symbol code is the same as in Figure \ref{Edotd2_P0}. When possible, the kinematic Shklovskii effect \cite{Shklovskii} has been taken into account in the calculation of $B_{LC}$ values. Globular cluster pulsars are not represented in this plot.\label{Blc_P0}}
\end{figure*}

The gamma-ray detectability does not seem to correlate with the value of the magnetic field at the stellar surface. Instead, the pulsars detected with the LAT have high values of the magnetic field at the light cylinder, $B_{LC} \propto \sqrt{\dot P / P^5} \propto \sqrt{\dot E / P^2}$. Figure \ref{Blc_P0} shows the values of $B_{LC}$ as a function of the rotational period $P$ for known pulsars. As we can see on this plot, both normal and millisecond pulsars detected in gamma rays are found in the upper part of their respective population. This again suggests that outer magnetospheric properties play an important role in gamma-ray emission for pulsars. For emission taking place in the outer magnetosphere, it is expected that the electric field parallel to the magnetic field lines is proportional to $B_{LC}$, in turn implying that the cutoff energy of the emitted spectra depends on $B_{LC}$ (see e.g. \cite{Muslimov2004,Zhang2004}). Such a correlation might indeed be observed (see Figure 7 of \cite{FermiPSRCat}), although strongly relying on the highest measured values of the cutoff energy $E_c$, which are rather uncertain. A reanalysis of these spectral parameters with increased datasets will reduce the uncertainties and make the relationship between cutoff energy and magnetic field at the light cylinder clearer.

\section{Search for off-pulse emission}

\begin{figure*}
\includegraphics[scale=0.5]{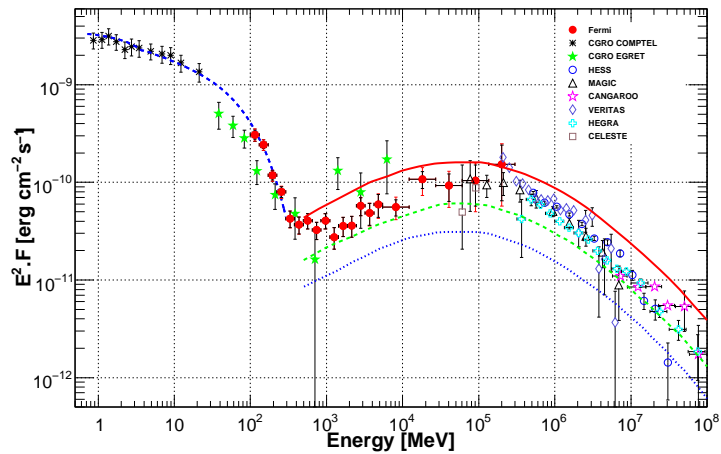}
\caption{Spectral energy distribution of the Crab Nebula, along with results of other experiments from soft to very high-energy gamma rays. See \cite{FermiCrabe} for details on the analysis, on the fits of the inverse Compton and synchrotron emission components and for references.\label{CrabNebula}}
\end{figure*}

Some young and energetic pulsars like the Crab are known to convert a large portion of their spin-down energy loss rate into a wind of energetic particles, the so-called ``Pulsar Wind Nebulae'' (PWNe), producing synchrotron and inverse-Compton emission. A number of these objects have been detected in the TeV domain, but the Crab Nebula was the only object ever detected in GeV gamma rays until \emph{Fermi} was launched. The off-pulse emission of the Crab pulsar, that is, the continuous emission outside of the pulsar gamma-ray peaks, has been analyzed using eight months of LAT data \cite{FermiCrabe}. The spectrum of the nebula is found to be well fit by the addition of a synchrotron and an inverse-Compton component, with no clear energy cutoff appearing for the synchrotron component. Figure \ref{CrabNebula} shows the spectral energy distribution of the Crab Nebula from soft to high-energy gamma rays. The spectrum measured with the LAT connects with lower and higher energy results within error bars, allowing for cross-calibration with Cherenkov experiments. 

\begin{figure*}
\includegraphics[scale=0.55]{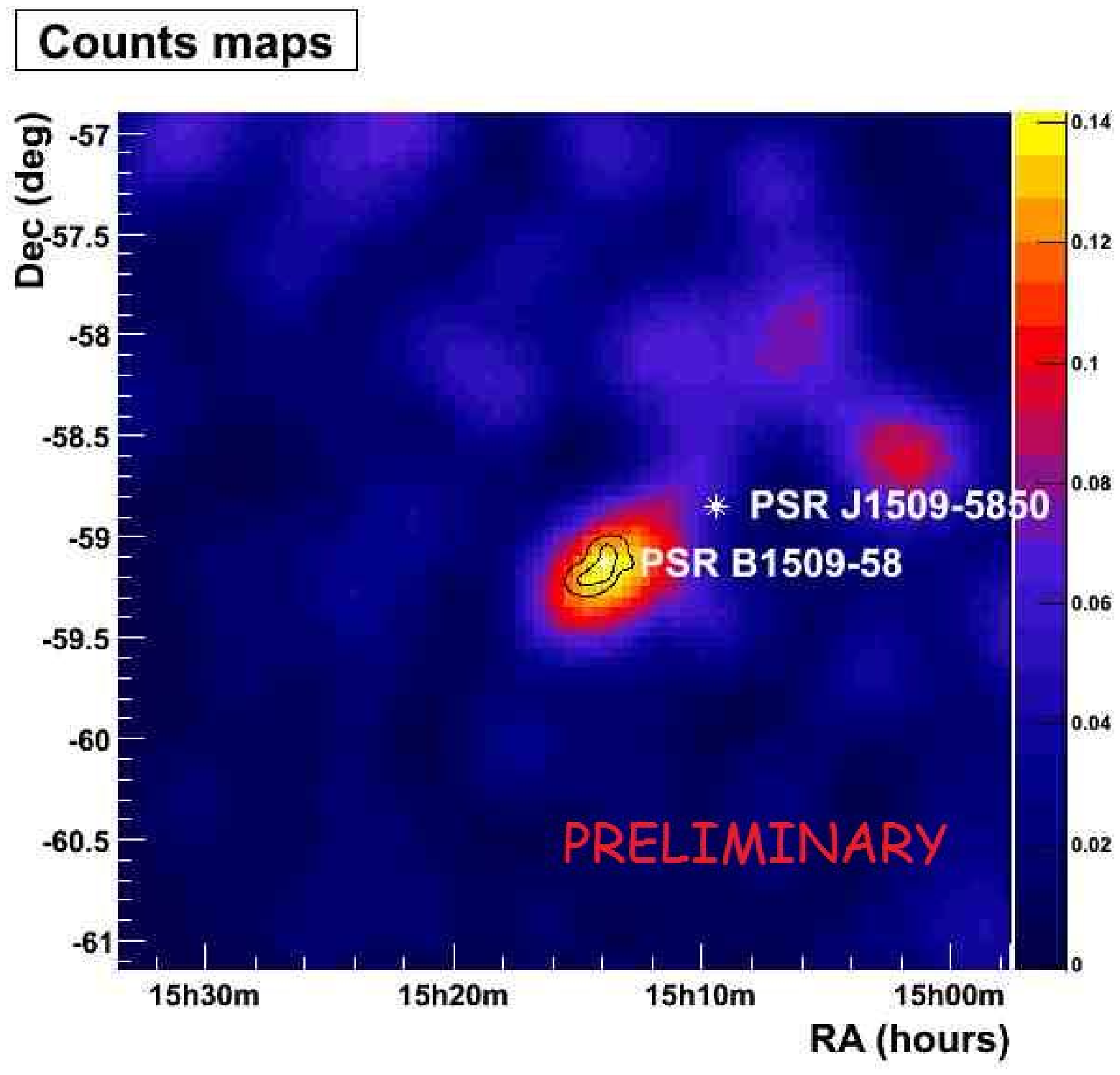}
\caption{Gamma-ray counts map around the MSH 15$-$52 pulsar wind nebula above 10 GeV, with H.E.S.S. contours overlaid. Also shown are the positions of the COMPTEL pulsar PSR B1509$-$58, and the PSR J1509$-$5850 detected with the \emph{Fermi} LAT. Figure adapted from \cite{MSH_MH}.\label{MSH_cmap}}
\end{figure*}

Besides the observation of the Crab Nebula with the LAT, almost half of the pulsars detected during the first year are found to be spatially associated with known X-ray or TeV PWNe and Supernova Remnants (SNRs) \cite{FermiPSRCat, FermiBlind2}. Some gamma-ray pulsars are associated with TeV sources not having counterparts at other wavelengths, which may point to yet unknown PWNe or SNRs. For instance, the association of the \emph{Fermi}-detected pulsar PSR J1907+0602 with the TeV source MGRO J1908+06 suggests that the latter is the pulsar wind nebula of the former \cite{FermiJ1907}. A systematic search for continuous (non-pulsed) emission in the off-pulse of gamma-ray pulsars is underway, and may foster detections of new PWNe or SNRs. Significant gamma-ray emission is already seen from objects other than the Crab nebula: the LAT detected GeV emission from the Vela-X pulsar wind nebula \cite{FermiVelaX}, providing constraints on the injected populations of charged particles. Another example is the detection of GeV gamma-ray emission from the pulsar wind nebula in MSH 15$-$52, powered by the COMPTEL pulsar PSR B1509$-$58 \cite{MSH_MH}. Figure \ref{MSH_cmap} shows a map of the observed emission above 10 GeV in one year of data. An excess is clearly observed around the PSR B1509$-$58 pulsar, coincident with H.E.S.S. contours of the pulsar wind nebula. Other such detections of nebular emission may be reported in the next months, and \emph{Fermi} LAT data along with multi-wavelength observations might help understand the physics of these objects associated with energetic pulsars. 

\section{Conclusion and prospects}

Pulsed gamma-ray emission has been observed for 55 pulsars over the first year, including previously unknown pulsars and millisecond pulsars. These pulsars are generally close and energetic (high $\dot E / d^2$ values), and have high magnetic fields at the light cylinder, $B_{LC}$. Their gamma-ray luminosity $L_\gamma$ increases with the spin-down energy loss rate $\dot E$. Low $\dot E$ pulsars seem to follow $L_\gamma \propto \dot E$, whereas high $\dot E$ ones rather follow $L_\gamma \propto \sqrt{\dot E}$. Refined spectral measurements and more pulsar detections will help see where the inflection occurs. Light curve shapes and spectra are comparable for normal and millisecond pulsars, indicating that the same mechanisms are operating. Gamma-ray pulsars seem to emit fan-like beams, separated from the radio emission (if any), and produced in the outer magnetosphere. Outer Gap and Slot Gap models are therefore favored over the Polar Cap model, although more data are required to discriminate between the two outer magnetospheric emission models. In addition, a few millisecond pulsars seem not to follow the Outer Gap and Slot Gap models, but are rather fit with a third family of high-altitude emission models, the Pair Starved Polar Cap model. Much has already been learned about the way gamma-ray pulsars emit; but unexpected features may appear in future pulsar observations, requiring refinement of current models. 

\begin{figure*}
\includegraphics[scale=0.67]{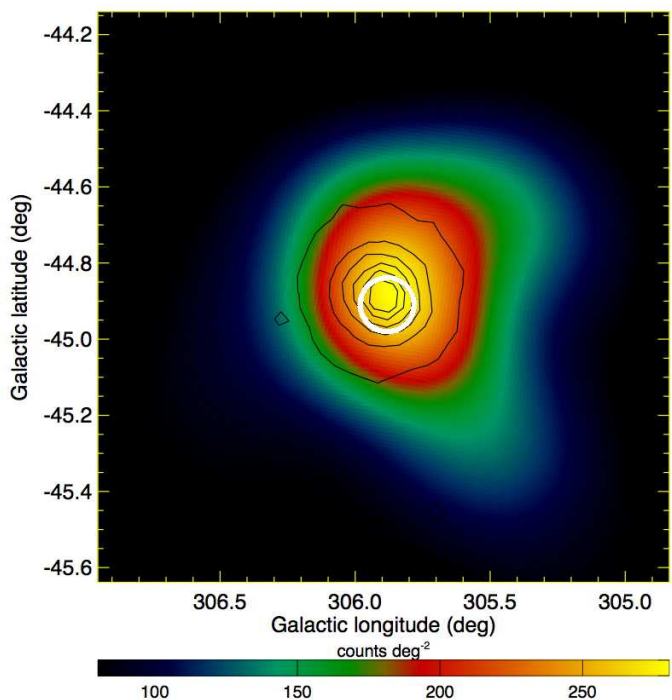}
\caption{$1.5^\circ \times 1.5^\circ$ region around the 47 Tucanae globular cluster, as seen with the LAT in eight months of data. Black contours show the stellar density in the cluster, while the white circle indicates the 95\% confidence region of the gamma-ray source. Additional details on this image can be found in \cite{Fermi47Tuc}.\label{47Tuc_cmap}}
\end{figure*}

The \emph{Fermi} LAT was initially expected to operate for five to ten years. If this goal is attained, then the LAT will eventually be two to three times more sensitive than it was during the first year. As can be seen in Figure \ref{Edotd2_P0}, the LAT has so far detected the highest $\dot E / d^2$ objects, and a lot of distant and/or low $\dot E$ pulsars may be awaiting detection, provided their gamma-ray beams are pointing towards the Earth. In addition to the simple accumulation of gamma-ray data, another way to increase the LAT sensitivity is to improve pulsation search algorithms. Such analysis refinements are being developed, increasing the sensitivity of the LAT at low energies or enhancing the pulsation search by giving more weight to photons that are likely associated with pulsars \cite{Burgess,Kerr}. 

More than half known MSPs are in globular clusters and are not represented in Figure \ref{Edotd2_P0}. None of these pulsars have yet been detected in gamma rays with the LAT, most probably because of their large distances. However, the LAT has detected continuous emission from the close globular cluster 47 Tucanae \cite{Fermi47Tuc} (see Figure \ref{47Tuc_cmap}). The observed spectra can be explained by the cumulative emission of seven to sixty-two MSPs. The LAT might eventually be able to see individual pulsars in globular clusters. Also, the blind searches of gamma-ray pulsars in LAT data have not been searching for unknown MSPs thus far. Some of these currently unknown MSPs may be radio-faint, and rapidly rotating radio-faint pulsars have never been detected, although they have all reasons to exist. 

Finally, there is much to expect from the follow-up observations of LAT unidentified sources at other wavelengths, such as those of the \emph{Fermi} LAT Bright Sources List \cite{FermiBSL}. First searches for pulsations at the position of LAT unidentified sources have already been extremely successful in finding new pulsars: for instance, several MSPs have been found at radio wavelengths with the Green Bank Telescope \cite{Ransom}. Similar searches are being conducted at Effelsberg, Nan\c cay, Jodrell Bank, Arecibo and Parkes radio telescopes. The discovered pulsars can then be monitored at radio wavelengths, and searched for gamma-ray pulsations in LAT data afterwards, using accurate rotation parameters. Radio telescopes have been leading pulsar science over the last 40 years, and the pulsars detected at optical, X-ray or gamma-ray wavelengths are still a minority compared to the $\sim$2000 observed in radio. For the first time though, LAT is providing radio telescopes with a nearly unbiased way to search for energetic normal and millisecond pulsars, by guiding them to unidentified gamma-ray sources in all directions of the sky. In addition to helping understand the physics of gamma-ray emission in pulsar magnetospheres, the LAT is also helping discover new pulsars at other wavelengths, some of them being of potential interest for gravitational wave detection projects, such as LEAP\footnote{http://epta.jb.man.ac.uk/leap.html} or NANOGrav\footnote{http://nanograv.org/}, or any project involving pulsars. We can therefore expect the \emph{Fermi} LAT to provide an important contribution to pulsar astronomy during the next few years.

\bigskip 
\begin{acknowledgments}

The \textit{Fermi} LAT Collaboration acknowledges generous ongoing support from a number of agencies and institutes that have supported both the development and the operation of the LAT as well as scientific data analysis. These include the National Aeronautics and Space Administration and the Department of Energy in the United States, the Commissariat \`a l'Energie Atomique and the Centre National de la Recherche Scientifique / Institut National de Physique Nucl\'eaire et de Physique des Particules in France, the Agenzia Spaziale Italiana and the Istituto Nazionale di Fisica Nucleare in Italy, the Ministry of Education, Culture, Sports, Science and Technology (MEXT), High Energy Accelerator Research Organization (KEK) and Japan Aerospace Exploration Agency (JAXA) in Japan, and the K.~A.~Wallenberg Foundation, the Swedish Research Council and the Swedish National Space Board in Sweden.

Additional support for science analysis during the operations phase is gratefully acknowledged from the Istituto Nazionale di Astrofisica in Italy and the Centre National d'\'Etudes Spatiales in France.\newline

\end{acknowledgments}

\bigskip 

\end{document}